\documentclass[a4paper]{jpconf}
\usepackage{graphicx}

\begin{document}

\title{Atomic Clock Ensemble in Space}

\author{L. Cacciapuoti$^1$, A. Busso$^1$, R. Jansen$^1$, S. Pataraia$^1$, T. Peignier$^1$, S. Weinberg$^1$, P. Crescence$^2$, A. Helm$^2$, J. Kehrer$^2$, S. Koller$^2$, R. Lachaud$^2$, T. Niedermaier$^2$, F.-X. Esnault$^3$, D. Massonnet$^3$, D. Goujon$^4$, J. Pittet$^4$, A. Perri$^4$, Q. Wang$^4$, S. Liu$^5$, W. Schaefer$^5$, T. Schwall$^5$, I. Prochazka$^6$, A. Schlicht$^7$, U. Schreiber$^7$, P. Laurent$^8$, M. Lilley$^8$, P. Wolf$^8$, C. Salomon$^9$}

\address{$^1$ European Space Agency, ESTEC, Noordwijk, The Netherlands}
\address{$^2$ Airbus Defence and Space, Friedrichshafen, Germany} 
\address{$^3$ CNES, Toulouse, France}
\address{$^4$ Safran Timing Technologies SA, Neuchâtel, Switzerland} 
\address{$^5$ Timetech, Stuttgart, Germany}
\address{$^6$ Czech Technical University in Prague, Prague, Czech Republic}
\address{$^7$ Technical University of Munich, Munich, Germany}
\address{$^8$ SYRTE, Observatoire de Paris-PSL, CNRS, Sorbonne Université, LNE, Paris, France}
\address{$^9$ Laboratoire Kastler Brossel, ENS-PSL, Sorbonne Université, Collège de France, CNRS, Paris France}

\ead{Luigi.Cacciapuoti@esa.int}

\begin{abstract}
The Atomic Clock Ensemble in Space (ACES) mission is developing high performance clocks and links for space to test Einstein's theory of general relativity. From the International Space Station, the ACES payload will distribute a clock signal with fractional frequency stability and accuracy of $1\times10^{-16}$ establishing a worldwide network to compare clocks in space and on the ground. ACES will provide an absolute measurement of Einstein's gravitational redshift, it will search for time variations of fundamental constants, contribute to test topological dark matter models, and perform Standard Model Extension tests. Moreover, the ground clocks connected to the ACES network will be compared over different continents and used to measure geopotential differences at the clock locations.

After solving some technical problems, the ACES flight model is now approaching its completion. System tests involving the laser-cooled Cs clock PHARAO, the active H-maser SHM and the on-board frequency comparator FCDP have measured the performance of the clock signal delivered by ACES. The ACES microwave link MWL is currently under test. The single-photon avalanche detector of the optical link ELT has been tested and will now be integrated in the ACES payload.

The ACES mission concept, its scientific objectives, and the recent test results are discussed here together with the major milestones that will lead us to the ACES launch.
\end{abstract}

\section{The ACES mission}
Atomic Clock Ensemble in Space \cite{Cacciapuoti2020, Laurent2015, Cacciapuoti2009} is a mission using high performance clocks in space and on the ground to test the foundations of general relativity, contribute to the generation and distribution of global time scales, and perform chronometric geodesy experiment.   

\begin{figure}
\begin{center}
\includegraphics[width=18pc]{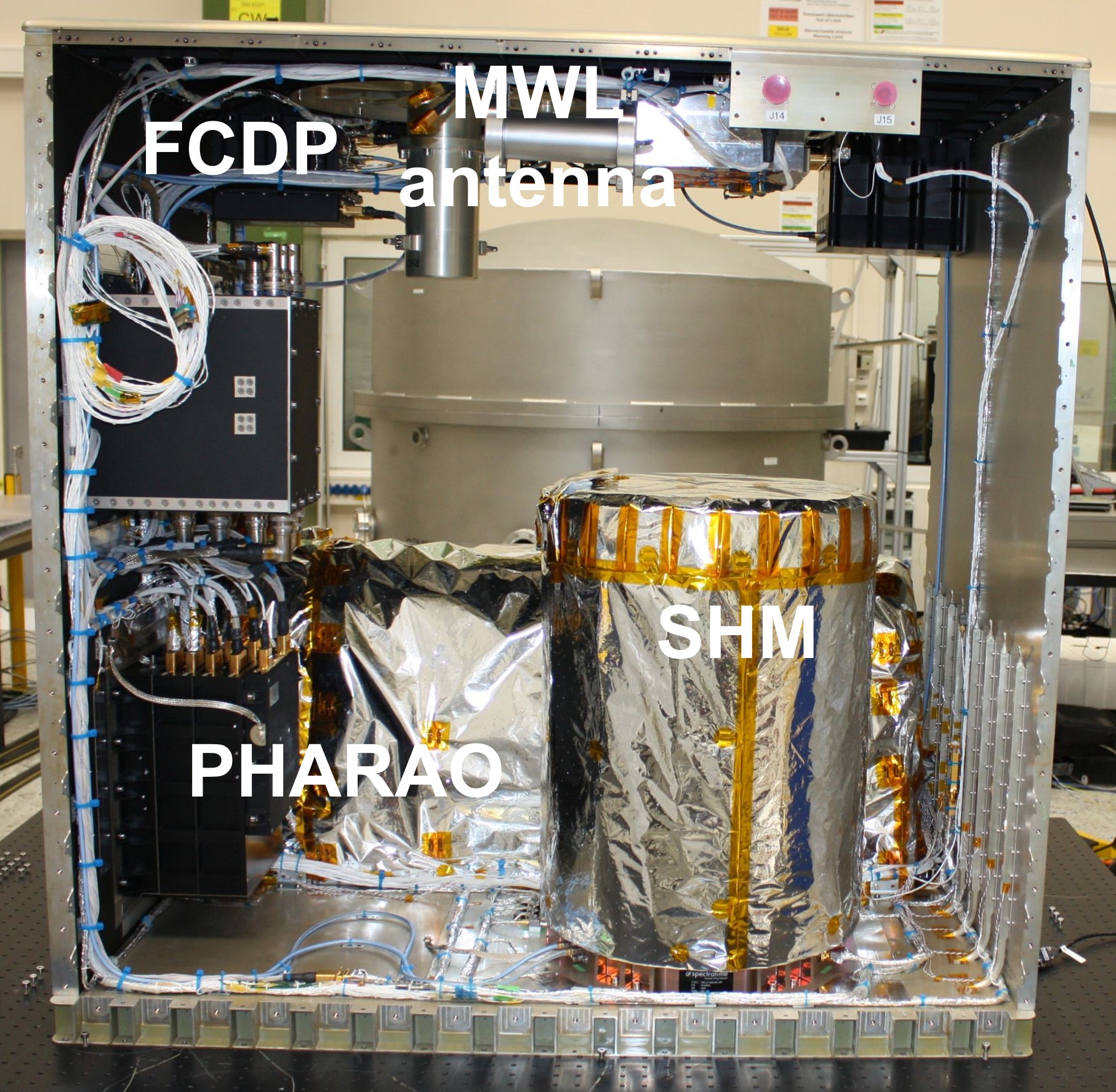}
\end{center}
\caption{\label{ACES}The ACES payload during integration. The vacuum chamber that will be used to test the payload is visible in the background.}
\end{figure}

The ACES payload (see Fig.~\ref{ACES}) will be installed onboard the International Space Station (ISS), at the external payload facility of the Columbus module. It includes two high stability and accuracy atomic clocks: PHARAO, acronym for Projet d'Horloge Atomique par Refroidissement d'Atomes en Orbite, is a primary frequency standard using laser-cooled Cs atoms; SHM, the ACES flywheel oscillator, is an active Space Hydrogen Maser. 

\begin{figure}[t]
\begin{center}
\includegraphics[width=38pc]{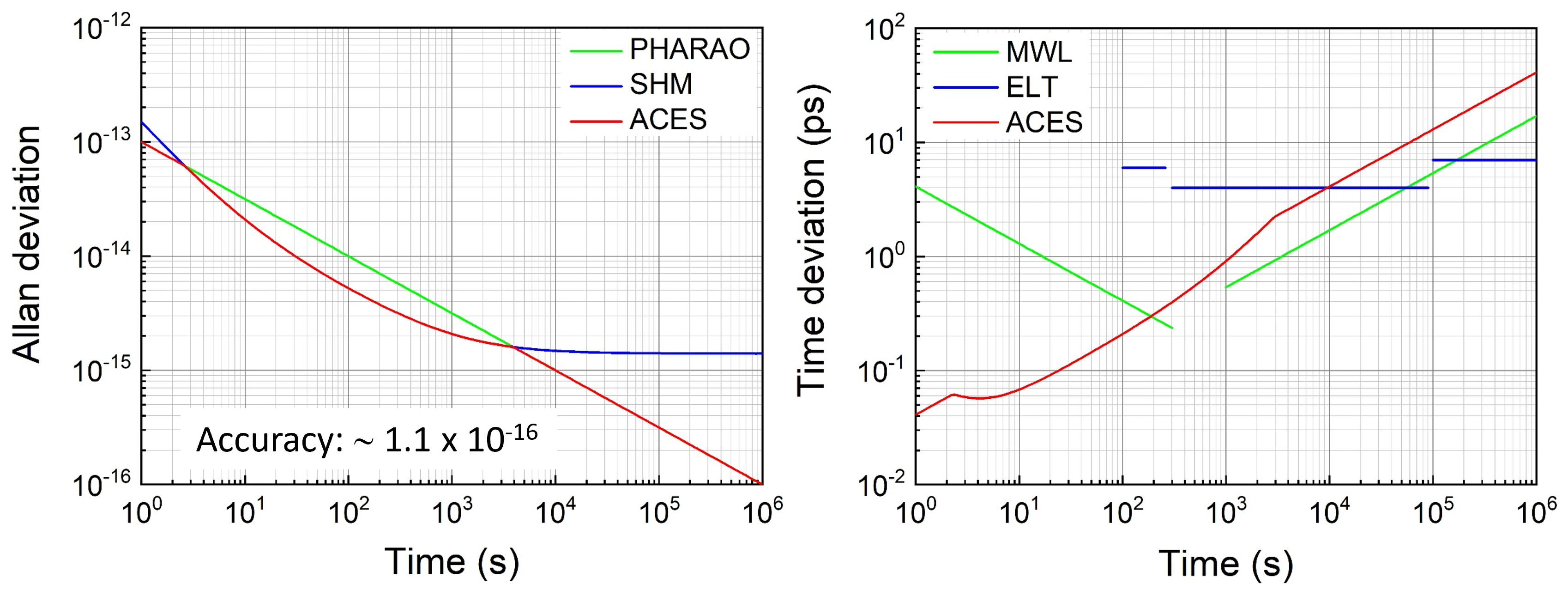}
\end{center}
\caption{\label{Perfo}(Left) Stability requirement of the ACES clocks expressed in Allan deviation. (Right) Time deviation specifying the stability of the ACES links for space-to-ground clock comparisons.}
\end{figure}

PHARAO~\cite{Laurent2006} is specified to a fractional frequency stability of $1\times10^{-13}/\sqrt{\tau}$ for $1~\textrm{s}<\tau<10^6~\textrm{s}$, where $\tau$ is the integration time expressed in seconds, and an accuracy of $1-2\times10^{-16}$  (see Fig.~\ref{Perfo}). The 100~MHz signals generated by PHARAO and SHM enter the Frequency Comparison and Distribution Package (FCDP), which compares the phase of the two clocks and distributes the ACES clock signal to the MicroWave Link (MWL) electronics. The phase comparison measurement between PHARAO and SHM is also the error signal of a phase locked loop (PLL) used to steer the local oscillator of PHARAO on the SHM clock signal on a time scale of 1-2~s (short-term servo-loop). A second loop, a frequency locked loop operating on a time scale of 100~s or longer (long-term servo-loop), will then use the frequency measurements performed in the PHARAO clock with respect to the Cs reference to correct SHM output against temperature variations and long-term drifts. The two loops lock the onboard clocks together thus defining the ACES frequency reference, which will have the short-term stability of SHM and the long-term stability and accuracy of the Cs clock PHARAO. At this point, the clock signal enters the MWL electronics where the ACES atomic time scale is built and used for the operation of the two onboard links. MWL is a 2-way, 3-frequency link \cite{ Meynadier2018, Hess2011, Duchaine2009}, with an up-link in the Ku-band and two down-links in the Ku and S-band. Here, the 100 MHz clock signal is upconverted, modulated by a PN-code carrying the time information, and sent towards ground users via the two antennas placed on the top panel of the ACES payload (see Fig.~\ref{ACES}) and facing the Earth. A distributed network of MWL ground terminals located at prominent metrology institutes worldwide (SYRTE, PTB, Wettzell Observatory, and NPL in Europe; JPL and NIST in the United States; NICT in Japan) will be the physical interface to connect ground clocks to ACES. Space-to-ground and ground-to-ground comparisons of clocks will be enabled by ACES. Moreover, a transportable terminal will be available to calibrate fixed MWL ground stations for time transfer experiments and to perform measurement campaigns on request at other metrology institutes. MWL is also responsible for time tagging the time-of-arrival of the laser pulses received by the ELT (European Laser Timing) optical link. A network of ground-based Satellite Laser Ranging (SLR) stations connected to atomic clocks is completing the ground segment needed for the ELT experiment. SLR stations will be coordinated by the Technical University of Munich within the International Laser Ranging Service (ILRS), to which ACES has already been proposed as an official ranging target. Wettzell SLR is the ACES primary station, which, together with Graz, Herstmonceux, Potsdam and Zimmerwald, has already been calibrated for time transfer experiments with ELT. Other stations can join, provided that they comply with the ISS safety standards. Finally, a dedicated GNSS receiver will provide precise orbit data of the ACES payload, important to operate the ACES time \& frequency links and precisely locate the PHARAO clock to perform redshift measurements.

The space-to-ground clock de-synchronization measurements delivered by MWL and ELT are the primary data product of the ACES mission. They are used to compare clocks both space-to-ground and ground-to-ground to test Einstein's theory of general relativity, and develop applications in time \& frequency metrology, and geodesy. 

In the fundamental physics domain, ACES will perform a gravitational redshift test on the Earth-to-ISS baseline. Differently from other redshift tests, the modulation of the redshift effect between perigee and apogee is quite small due to the reduced eccentricity of the ISS orbit and therefore marginally sufficient for a competitive measurement. Therefore, ACES will rely on the accuracy of the PHARAO clock and of Cs fountain clocks on the ground to perform an absolute measurement of the frequency difference between the space and the ground clock. The ACES performance for a redshift test along the ISS orbit has been studied in \cite{Savalle2019}. With PHARAO uncertainty at $1-2\times10^{-16}$ and a network of up to 8 ground stations, an uncertainty of 2-3~ppm is expected to be reached after 20 days of measurement, a factor 70 improvement on the historical GPA experiment \cite{Vessot1980, Vessot1979}, and a factor 10 better than the more recent redshift test based on the GALILEO 5 and 6 satellites \cite{Delva2015, Herrmann2015}. The characterization of the PHARAO accuracy budget to $1-2\times10^{-16}$ will require the complete mission duration to minimize the statistical uncertainty on the evaluation of the frequency shifts due to cold collisions and distributed cavity phase. 

\begin{figure}[t]
\begin{center}
\includegraphics[width=28pc]{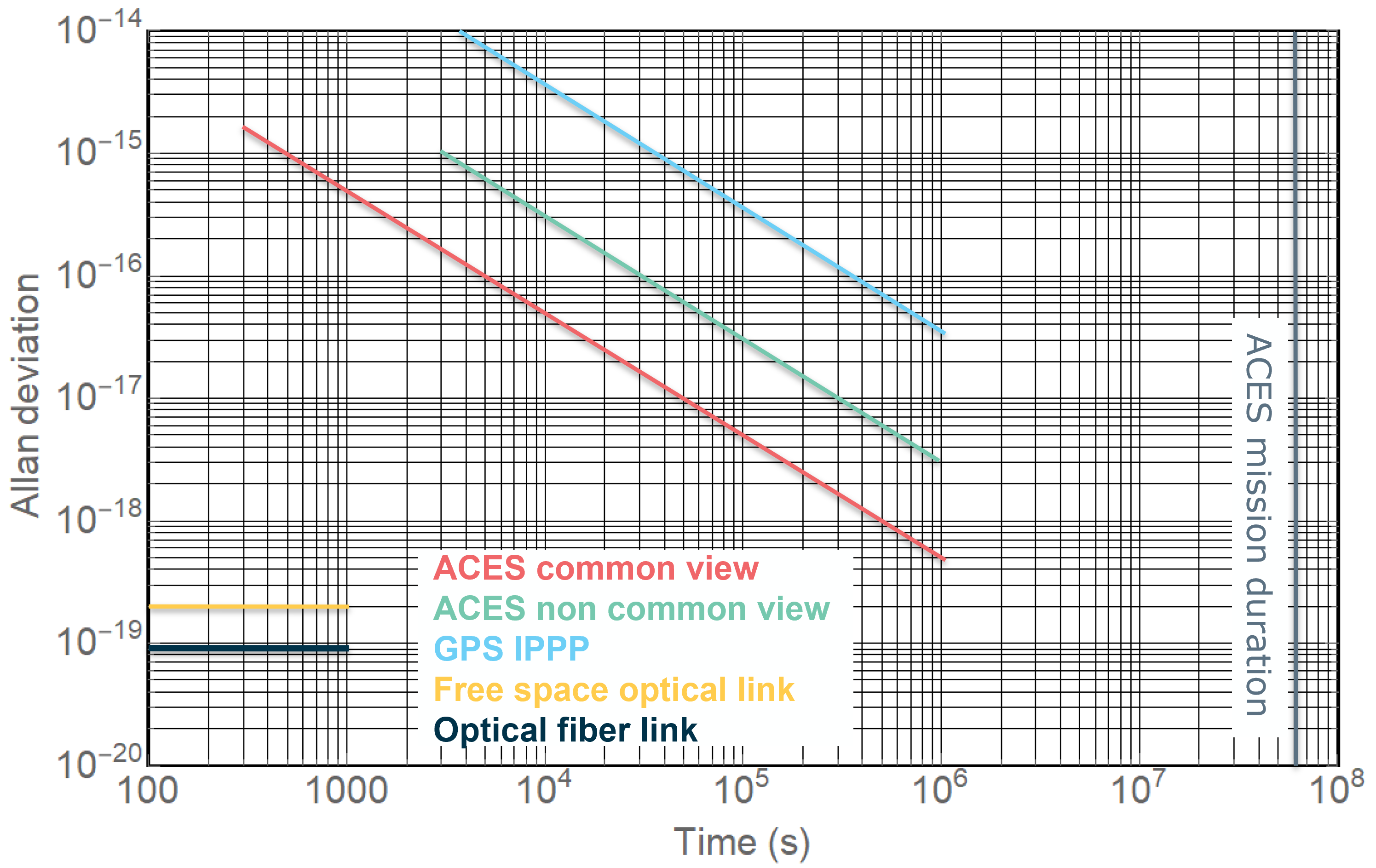}
\end{center}
\caption{\label{G2G}ACES MWL performance for the comparison of two ground clocks in common view (red) and non-common view (green)~\cite{Cacciapuoti2009} compared to the GPS IPPP performance (blue) \cite{Petit2017}. The typical stability of optical links through a fiber (black) or in free space (yellow) are also reported \cite{Dix-Matthews2021}.}
\end{figure}

One of the most important aspects of ACES is its network potential. The stability specified for the ACES links for space-to-ground clock comparisons is reported in Fig.~\ref{Perfo}. However, ACES can also be seen as a relay satellite connecting clocks in space and on the ground in a worldwide network. Ground clocks can be compared when in common view with the ACES payload, or across continents in the non-common view configuration. In the first case, the noise of the onboard clock is in common-mode and cancels when taking the difference of the two simultaneous space-to-ground comparisons. As shown in Sec.~3, both MWL and ELT long-term stability remains below 1~ps. Therefore, the performance of the common view comparison, now depending on the link noise only, allows reaching a fractional frequency uncertainty of $1\times10^{-17}$ after a few days of integration time. In a non-common view comparison, the SHM clock will act as a flywheel oscillator keeping the time across continents. The uncertainty in the comparison of the two ground clocks will now depend on both the link noise and the time error accumulated by SHM during the deadtime. Due to the extra noise introduced by the clock error, non-common view comparisons will require about 5 days to reach the $1\times10^{-17}$ uncertainty level. Figure~\ref{G2G} shows the ACES performance for common view (red) and non-common view (green) comparisons of two ground clocks~\cite{Cacciapuoti2009}, together with the GPS IPPP performance (blue)~\cite{Petit2017} and the typical stability of optical links though a fiber (black) and in free space (yellow) \cite{Dix-Matthews2021}.

Optical atomic clocks have experienced a tremendous progress in the last decades \cite{Nicholson2015, McGrew2018, Brewer2019, Bothwell2019, Takamoto2020} and ACES will be able to exploit their potential connecting them in a worldwide network. Cross comparisons of atomic clocks based on different transitions will allow testing the time stability of the fine structure constant, the quark mass, and the electron mass \cite{Dzuba1999a, Dzuba1999b}. In a similar way, the ACES network of atomic clocks will also be used to search for dark matter and place boundaries on topological dark matter models~\cite{Stadnik2015a, Stadnik2015b, Derevianko2014}. Topological dark matter can be represented by a scalar field that couples to the three fundamental constants thus inducing fluctuations of atomic transition frequencies. The comparison of clocks based on different atomic transitions will allow to place limits on the three energy scales, which are defining the coupling of the dark matter field to the three fundamental interactions. The simultaneous observation with several clocks compared along different baselines will provide ways to confirm any observation above the sensitivity threshold and control systematic errors. Moreover, the screening effect of the dark matter field due to the Earth mass will be reduced to about 0.06 on the space clock PHARAO, several order of magnitude less than the $10^{-7}$ shielding factor experienced by ground clocks \cite{Schkolnik2023}. Therefore, the PHARAO clock, which is sensitive to all three fundamental constants, will assume a key role in the hunt for dark matter.

The ACES network of ground clocks will also be used for chronometric geodesy experiments \cite{Takamoto2020, Takano2016, Grotti2018, Lion2017} on a worldwide scale. The measurement performed at the Tokyo Skytree Tower has demonstrated the possibility of resolving the geopotential difference between two transportable optical lattice clocks at the level of $5\times10^{-18}$ in relative frequency \cite{Takamoto2020}. ACES intercontinental comparisons of atomic clocks will extend this method to a global scale by providing a resolution on the geoid height better than 10~cm. The coverage offered by ACES will thus complement the results of the CHAMP, GRACE, and GOCE missions providing fixes with a very high spatial resolution to constrain existing Earth gravity models.

In the time \& frequency metrology domain, the ACES microwave and optical links will also be used to synchronize clocks on the ground. While MWL will provide 100~ps accuracy in ground-to-ground time transfer experiments, the optical link ELT is expected to reach even better performance, down to the 50~ps level. Finally, by connecting optical clocks over four continents to a fractional frequency uncertainty of $1\times10^{-17}$ or better, the ACES links will help the process towards the realization of a new definition of the SI second based on an optical frequency standard. Optical clocks and space links are clearly the next generation technology for the distribution of time and frequency on a worldwide scale.

The ACES mission will be commanded and controlled from the ACES Users Support and Operations (USOC) center, which will be located in CADMOS, at the CNES campus in Toulouse (FR). From there, we will be able to send commands to the ACES payload and to the MWL ground terminals, and to receive the science and housekeeping telemetry. CADMOS is also the central hub where all the ACES data will be stored and analysed to generate quick-look products for the monitoring and control of the flying instruments. From there, the data will be accessed by the data analysis centers and further downstream by the scientific community. SYRTE is hosting the MWL data analysis center and is responsible for providing full performance space-to-ground clock de-synchronization measurements and higher level data products. The Technical University of Munich will be the only interface between the ACES USOC and SLR stations contributing to the ELT experiment, also responsible for the ELT data analysis. Finally, orbitography products will be generated by the German Space Operations Center (GSOC), which is located in Oberpfaffenhofen (DE).

ACES will be launched to the ISS on a Space X Dragon vehicle carried by a Falcon 9 rocket. After docking to the ISS, the robotic arm will extract the ACES payload from the Dragon trunk and install it at the lower balcony of the Columbus module, with the MWL antennas looking towards the Earth. The first 6 months on orbit will be dedicated to commissioning activities, which include switch-on, temperature control optimization, instruments tuning, and calibrations. During this phase, the performance of the ACES clock signal and of the time \& frequency links will be optimized in preparation for the routine science phase. ACES will then be operated in nominal mode for the next 2 years, during which the PHARAO systematic frequency shifts will be measured with increasing precision and the links continuously used to compare the ACES clock signal to ground clocks.  

\section{The ACES clock signal}
The ACES clock signal combines the medium-term frequency stability of SHM with the long-term stability and accuracy of the PHARAO clock.

\begin{figure}[t]
\begin{center}
\includegraphics[width=24.5pc]{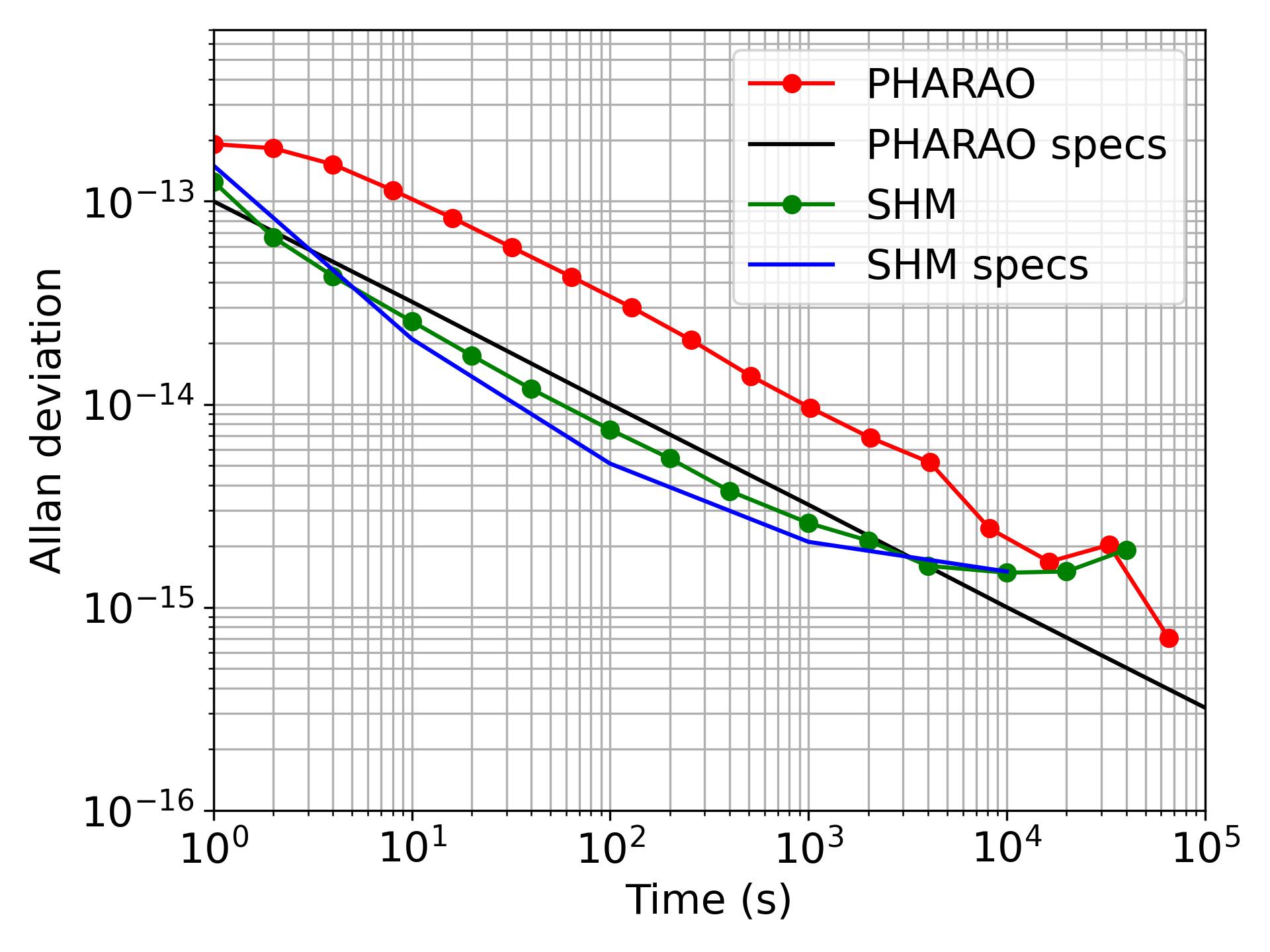}
\end{center}
\caption{\label{PHARAO_SHM}Fractional frequency stability of PHARAO in autonomous mode and of SHM in stand-alone configuration during ground tests. In microgravity, the 5.6~Hz linewidth of the PHARAO resonance will be reduced by one order of magnitude, enabling the specified stability indicated by the black line~\cite{Laurent2020}. The plots include the noise contribution from the reference H-maser. After removing it, SHM stability becomes compatible with the specified performance.}
\end{figure}

PHARAO is a primary frequency standard designed to operate under weightlessness conditions \cite{Laurent2006}. Cs atoms are laser cooled in an optical molasses to a temperature of $1-2~\mu$K. The cooling process produces a cloud of 15-20~mm diameter containing $5\times 10^8$ atoms after a loading time of about 1~s. Using the moving molasses technique, the atomic cloud is launched along the PHARAO tube where it crosses two microwave cavities in resonance with the clock transition at 9.192631770~GHz (definition of SI-second). In the first microwave cavity, the atoms are prepared in the initial clock state; in the second, they are interrogated on the clock transition in a Ramsey sequence. At the end of the tube, the atomic population in the two hyperfine levels of the Cs ground state is selectively measured by fluorescence detection. The cooling and loading process, the launch, the preparation, the interrogation, and detection phases have been optimized. Figure~\ref{PHARAO_SHM} shows the fractional frequency stability of the PHARAO clock when operated in autonomous mode. 
In this configuration, the frequency of the 100~MHz oscillator of PHARAO is continuously measured against the Cs frequency reference and used in a feedback loop to stabilize the 100~MHz clock output. On the ground, Cs atoms are launched against gravity with a velocity of 3.54~m/s. The limited transit time between the two interrogation zones of the PHARAO Ramsey cavity defines the width of the central fringe of the Ramsey resonance, which can be 5.6~Hz, limiting the fractional frequency stability of the clock to $3.3\times10^{-13}/\sqrt{\tau}$, where $\tau$ is the integration time expressed in seconds. In space, under free fall conditions, Cs atoms will be launched very slowly and, thanks to a narrower resonance of 0.2-0.5~Hz, a frequency stability of $1.1\times10^{-13}/\sqrt{\tau}$ will be reached. The frequency accuracy of the PHARAO clock has been evaluated several times in the past years \cite{Laurent2006, Laurent2020}. The PHARAO accuracy budget, measured on the ground to $1-2\times10^{-15}$, has also been checked by performing an absolute frequency measurement between the PHARAO clock at Airbus premises in Friedrichshafen (DE) and the Cs fountain clocks operated by SYRTE in Paris (FR) through a GPS link. The frequency measurement was found compatible with zero within $1.5\times10^{-15}$. The major contributions to the PHARAO accuracy on the ground arise from the collisions between cold atoms and the distributed phase of the microwave interrogation field. The distributed cavity phase shift is strongly increased by the power imbalance between the two pulses of the Ramsey cavity due to the gravity deceleration. In space, with a narrower linewidth, a much lower atomic density, and a well balanced Ramsey interaction, it will be possible to characterize PHARAO accuracy to $1-2\times10^{-16}$. PHARAO has completed its acceptance test and it is integrated in the ACES payload.   

SHM is an active H-maser designed for space. Despite the significantly reduced mass, volume, and power consumption (46~kg, 70~l, 93~W), SHM frequency stability is very close to the one of active H-masers on the ground. SHM has a crucial role on the ACES payload. It is the stable frequency reference used in space to characterize the PHARAO systematic frequency shifts and determine its accuracy budget. Moreover, it is the ACES flywheel oscillator that, distributed by FCDP, drives the MWL electronics, where the ACES atomic time scale is built to be sent to ground users. After the failure of the ion pump, SHM went through a major refurbishment activity to replace the getter material and activate it. SHM has now completed its acceptance tests and it has recently been integrated in the ACES payload. Its frequency stability has been characterized against a reference H-maser in terms of Allan deviation (see Fig.~\ref{PHARAO_SHM}). After removing the contribution of the reference H-maser the clock stability becomes $7.9\times10^{-14}$ at 1~s of integration time, $2.0\times10^{-14}$ at 10~s, $6.2\times10^{-15}$ at 100~s, $2.2\times10^{-15}$ at 1000~s, reaching the $1\times10^{-15}$ level at $10^4$~s, compatible with the Allan deviation specified for SHM. 

\begin{figure}[h]
\begin{center}
\includegraphics[width=24.5pc]{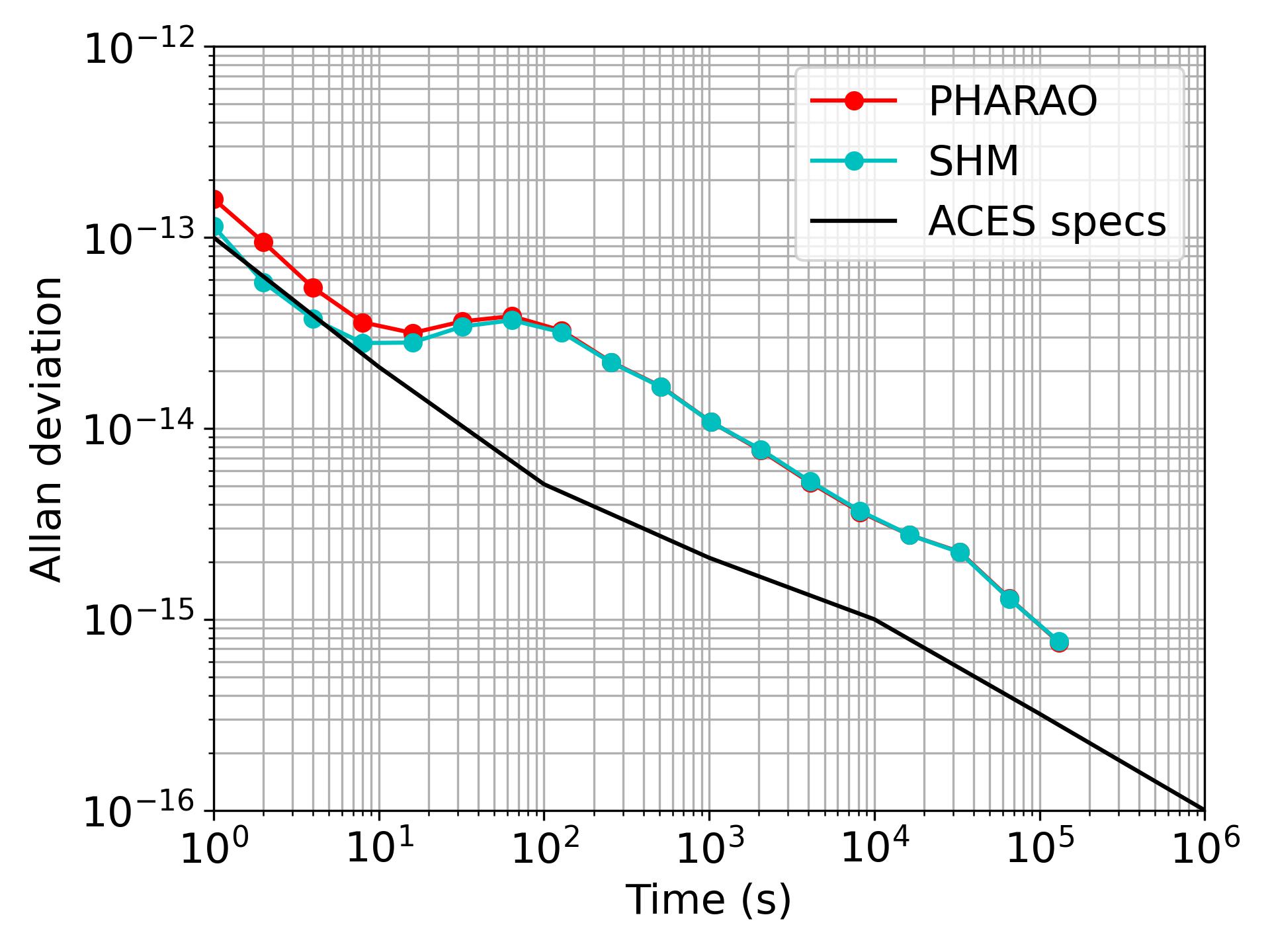}
\end{center}
\caption{\label{ACESClock}Fractional frequency stability of PHARAO and SHM when the short-term and long-term servo loops are closed. After 100~s, the long-term servo loop forces the SHM clock signal to follow PHARAO stability curve (see also Fig.~\ref{PHARAO_SHM}).}
\end{figure}

The ACES clock signal is generated from PHARAO and SHM after closing the ACES servo-loops. In this configuration, the 100~MHz clock signal from SHM has the short-term to medium-term frequency stability of the active H-maser and the long-term stability and accuracy of the Cs clock PHARAO. The Allan deviation of SHM and PHARAO when the ACES servo-loops are closed is shown in Fig.~\ref{ACESClock}. The two curves overlap to a large extent after 40~s of integration time. The short-term servo loop forces PHARAO to follow the SHM clock. The discrepancy that can be observed between PHARAO and SHM below 40~s is due to the noise of the FCDP phase comparator, which is transferred to the PHARAO 100~MHz clock signal by the short-term servo-loop. This noise averages down as $1/\tau$, where $\tau$ is the integration time, and it becomes negligible already after 40~s. For integration times longer than 100~s, the long-term servo loop forces the SHM clock signal to follow the PHARAO stability curve (see also Fig.~\ref{PHARAO_SHM}).

The ACES servo-loops are further optimized during the final system tests, which involve the complete ACES payload. During this campaign, another mode of operation, called the PHARAO evaluation mode, will also be tested. In the evaluation mode, the long-term servo-loop is kept open with the PHARAO clock continuously measuring the frequency offset between SHM and the Cs resonance. This configuration avoids any frequency offsets or instabilities that might be introduced by the long-term servo-loop on the PHARAO clock and allows to characterize any perturbations injected on the PHARAO local oscillator through the short-term servo-loop (e.g. AM/PM conversion effects at the ACES phase comparator FCDP). With PHARAO in evaluation mode, the SHM clock signal distributed by ACES to ground users is freely running and the long-term stability and accuracy of the PHARAO clock are not available in real-time. However, they can be retrieved in post-processing from the measurements of the frequency difference between PHARAO and SHM provided in the telemetry of the PHARAO clock.   

\section{The ACES time \& frequency transfer links}

MWL is a 2-way 3-frequency link in the microwave domain \cite{ Meynadier2018, Hess2011, Duchaine2009}. The 2-way link in the Ku-band is used to efficiently remove the tropospheric time delay and the first-order Doppler effect, and finally to measure the space-to-ground clock de-synchronization. In addition, the symmetry between space and ground transceiver will allow reducing phase perturbations introduced by the signal dynamics (e.g. tracking loop errors or AM to PM conversion effects). The second downlink in the S-band is used with the Ku-band downlink to determine the Total Electron Content (TEC) of the ionosphere and correct the ionospheric time delay. The phase of the Ku-band carrier is modulated (BPSK) by a Pseudo-Random Noise (PN) code at 100~MChip/s. This modulation, coherently generated from the ACES clock, superimposes unambiguous time tags on the carrier signal, thus defining the ACES time scale. The high chip rate ensures high resolution code phase measurements, which are important to identify the cycle of the carrier. In addition it helps reducing multipath effects. 4 receiver channels are able to simultaneously compare the ACES clock signal with 4 atomic clocks on the ground. In each receiver channel, a Delay Locked Loop (DLL) acquires the PN code and tracks it. The carrier of the incoming signal is extracted after the correlator and tracked in a PLL with respect to the local replica. Every 80~ms, the space and ground transceivers provide a code and carrier phase measurement, which determines the time difference between a given transition of the PN code and zero crossing of the carrier in the received signal and the same code transition and carrier zero crossing in the signal locally generated by the transceiver, coherently with the local clock. From the combination of the phase measurements performed by the MWL transceivers in space and on the ground, we extract the MWL scientific products, which include the space-to-ground clock de-synchronization, the Total Electron Content (TEC) of the ionosphere, and the combination of the round trip pseudorange and tropospheric delay.   

\begin{figure}
\begin{center}
\includegraphics[width=24pc]{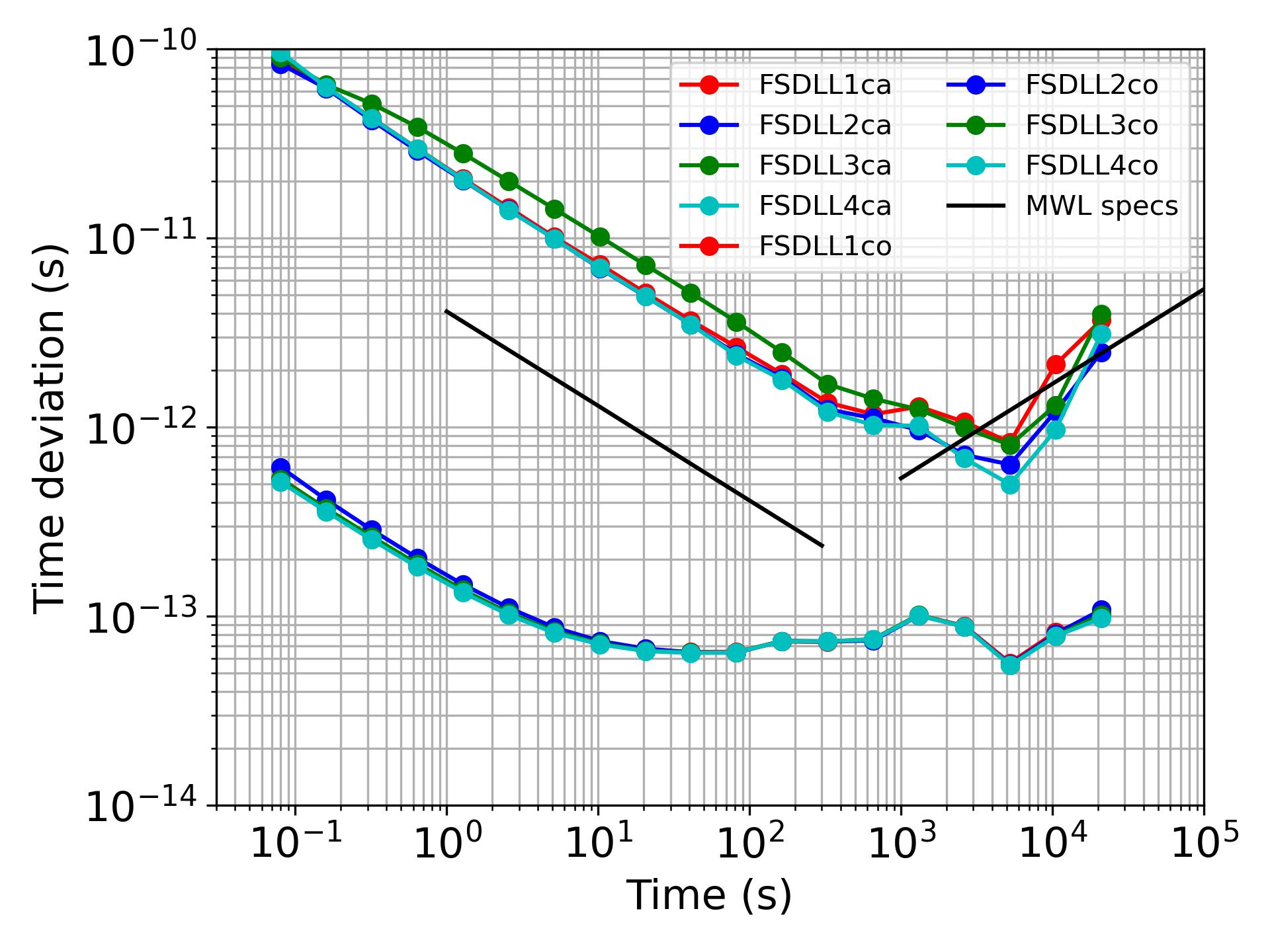}
\end{center}
\caption{\label{MWL_SE2E} Time stability of code (co) and carrier (ca) phase measurements for the 4 DLL receivers of the MWL flight segment electronics.}
\end{figure}

To test the stability of code and carrier phase measurements, we connect the MWL flight transceiver to the ground terminal one using cables. Both transceiver are connected to the same clock signal (common-clock test) and are operated under static conditions with a transmit signal of -108~dBm, corresponding to the power level expected at reception when the ISS reaches its maximum elevation above a ground station in Europe. Figure~\ref{MWL_SE2E} shows the stability of code and carrier phase measurement expressed in time deviation. Code phase measurements reach a sub-ps noise floor after about 1000~s of integration time. With such performance, the cycle of the carrier can be resolved in less than 10 s of integration time. However, the ultimate stability is obtained by using carrier phase measurements, which reach a time deviation of 60~fs after 10~s and remain well below 1~ps on the long term. 

\begin{figure}
\begin{center}
\includegraphics[width=38pc]{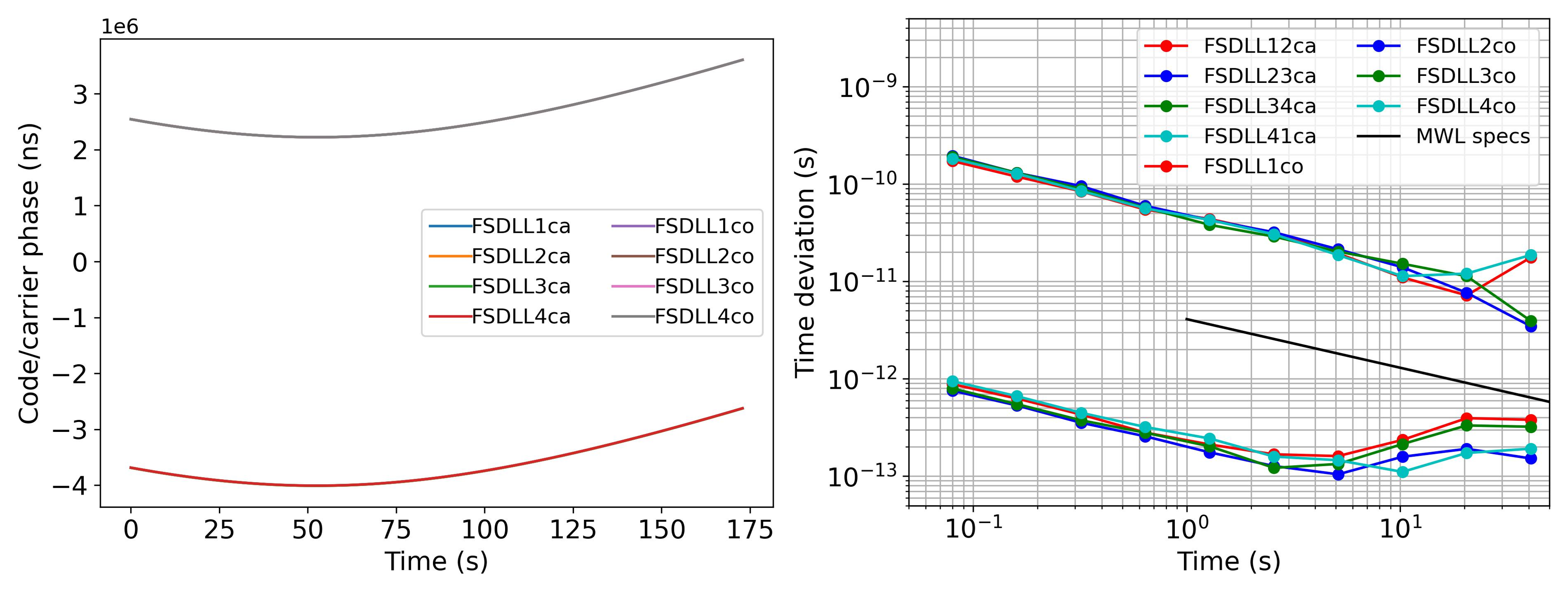}
\end{center}
\caption{\label{MWL_DE2E}(Left) Evolution of code and carrier phase measurement during the simulated ISS pass. (Right) Time stability of the difference of code (co) and carrier (ca) phase measurements for the 4 DLLs of MWL flight segment electronics.}
\end{figure}

MWL tests in the presence of signal dynamics are currently ongoing. In this case, the MWL flight segment transceiver is connected to the ground terminal through an RF system that generates the Ku-band uplink, the Ku-band downlink, and the S-band downlink signals with the dynamical behavior (in Doppler frequency and amplitude) expected along the ISS orbit. Such tests are important to determine the link sensitivity to AM/PM conversion effects, group delays, and tracking loop error. 
To evaluate the flight transceiver performance in the presence of signal dynamics, we repeatedly execute three passes with low (16$^{\circ}$), medium (36$^{\circ}$) and high elevation (84$^{\circ}$) in parallel for all 4 receiving DLLs. As the pass starts, MWL DLLs enter the lock acquisition mode. In this phase, the frequency of the code and carrier oscillators is steered in open loop to follow the incoming signal. The signal is correlated in the DLL against the local replica of the PN-code while the range is scanned, searching for the correlation peak. At this point the delay locked loop is closed and the code oscillator tracks the frequency variations of the incoming signal. After acquisition of the 1PPS (Pulse Per Second) marker, also the PLL on the carrier is closed. At this point, both code and carrier phase measurements are correctly tracking the incoming signal. The 3-pass sequence was repeated several times for a total of 200 passes. All 4 receiver channels acquired lock correctly, albeit with different performance. DLL 1 and 3 entered the full tracking mode in less than 50~s, corresponding to an elevation of 10$^\circ$. Closing the loop on the incoming signal with DLL 2 and 4 takes longer, requiring further optimization of the loop parameters. Figure~\ref{MWL_DE2E} (left) shows the typical profile of code and carrier phase measurements during a pass. The code phase is non-ambiguous and results from the sum of the integrated Doppler frequency of the incoming signal and the initial de-synchronization between the MWL flight electronics and the RF system generating the dynamic signals. On the contrary, carrier phase measurements can only be resolved up to a constant offset after determining the integer number of carrier cycles with respect to code phase measurements. The large phase variations due to the integrated Doppler frequency can be cancelled by calculating the phase difference between two DLLs. Figure~\ref{MWL_DE2E} (right) shows the stability of code and carrier phase differences between the 4 DLLs of the MWL flight transceiver. After removal of the phase variations due to the Doppler effect, the time stability reaches levels compatible with the flicker floor measured during static tests (see Fig.~\ref{MWL_SE2E}). This analysis is useful to infer the performance of the end-to-end link in the presence of signal dynamics, however it does not reflect  the ultimate performance of MWL. Indeed, on the one hand, the phase difference between two DLLs removes the correlated instrumental noise of the receiving channels that would naturally enter the comparison between the space and the ground clock; on the other hand, signal dynamics effects like AM/PM conversion, tracking look errors, and group delay dependence on the Doppler frequency, which are DLL specific and need to be calibrated, are removed only partially.  

End-to-end tests in the presence of signal dynamics are continuing, together with the calibration of signal dynamics effects. In addition, the temperature sensitivity of code and carrier phase measurements, the sensitivity to supply voltage, and clock input power will be determined. Soon, all the elements to correctly combine the phase measurements from the flight and ground transceivers in the 2-way formula will be at hand and we will be able to assess the ultimate performance of MWL on ISS-representative passes.

ELT is an optical link that uses short laser pulses exchanged between a satellite laser ranging (SLR) station on the ground and ACES to determine the de-synchronization between ACES and the SLR station clock. The SLR station ranges the ISS by firing laser pulses with a duration of about 40-60~ps. The laser pulse emission event is time stamped on the ground in the local clock time scale ($t_{start}$). Once it reaches the ISS, the pulse is detected by the ELT Single-Photon Avalanche Detector (SPAD) and time stamped in the ACES time scale, which is generated in MWL ($t_{space}$). At the same time, the onboard corner cube reflector redirects the echo return to the SLR station where its time of arrival is recorded ($t_{stop}$). From the three time tags, it is possible to calculate the space-to-ground clock de-synchronization as
\begin{equation}
    \tau_{g}(t)-\tau_{s}(t)=\frac{t_{start}+t_{stop}}{2}-t_{space}+\tau_{rel}+\tau_{atm}++\tau_{geom}
\end{equation}
where $\tau_{g,s}(t)$ is the proper time on the ground and in space at coordinate time $t$, and $\tau_{rel}$, $\tau_{atm}$, and $\tau_{geom}$ are the delays due to relativistic effects, propagation in the atmosphere, and geometrical ties. 

\begin{figure}
\begin{center}
\includegraphics[width=25pc]{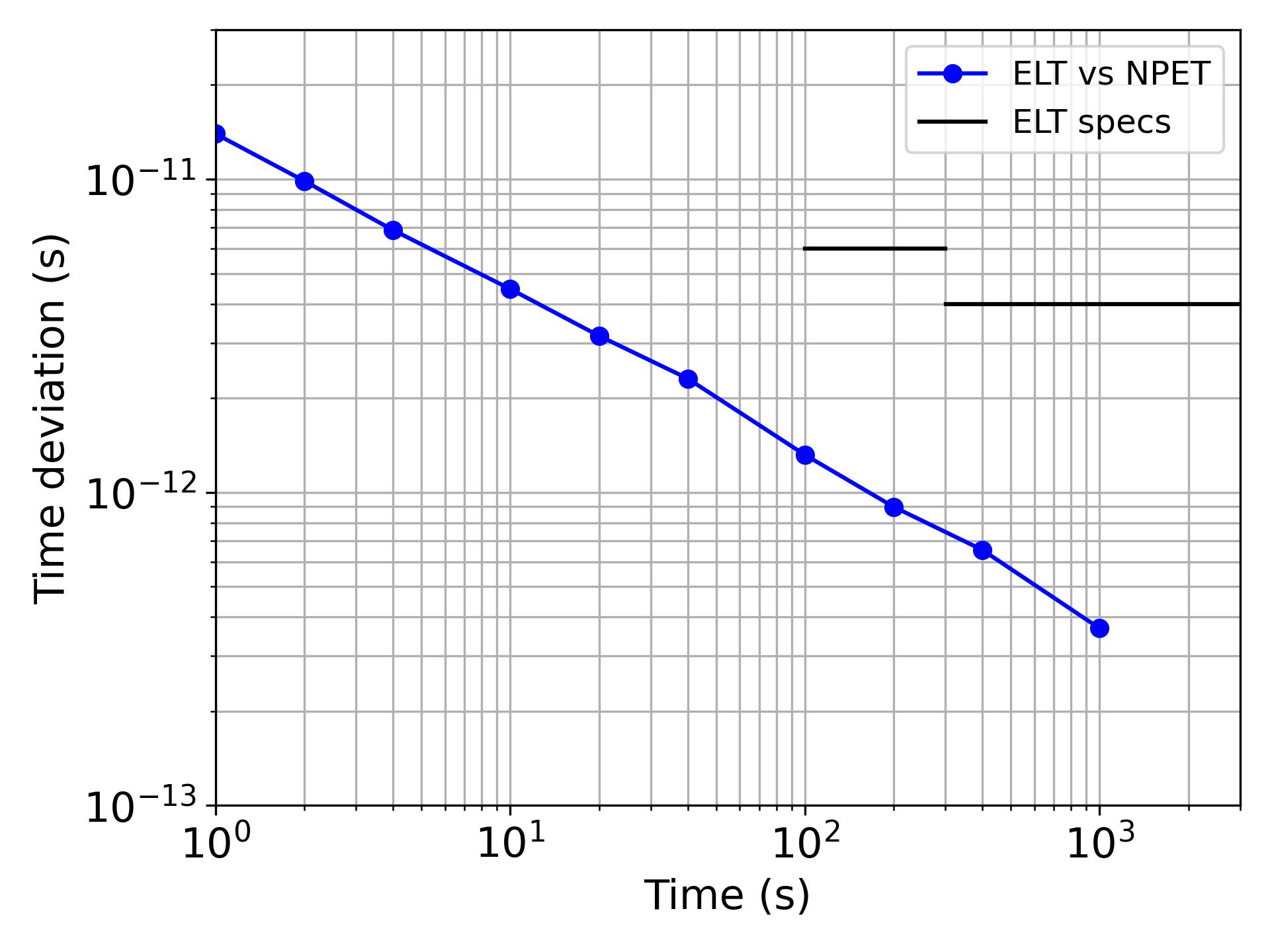}
\end{center}
\caption{\label{ELT}Time stability of the ELT detection chain. The SPAD is operated in dark count mode and the detection events are measured in parallel by the MWL time tagging board and an external event timer, both referenced to a common clock. In this experiment, the SPAD unit of ELT is gated with 100~Hz repetition rate.}
\end{figure}

The ELT detection chain comprises the SPAD unit and the electronics for time tagging the detection events, which is integrated in MWL \cite{Schreiber2010}. The temperature stability of the detector package was tested between -55$^{\circ}$C to +60$^{\circ}$C, showing a temperature sensitivity of 1~ps/K over the entire temperature range. The optical-to-electrical detection delay was also characterized with an uncertainty of 20~ps \cite{Prochazka2011}. To test the stability of the ELT flight hardware, we connect the SPAD flight unit in parallel to MWL and to an external time tagging unit \cite{Prochazka2011}. The time scales of both MWL and the external event timer are referenced to the same clock signal. With the SPAD aperture closed by a light-blocking cap and the detector operated in dark count mode, we measure the difference between the detection events recorded in MWL and the external event timer. Figure~\ref{ELT} shows the time deviation of the time differences obtained by gating the SPAD unit with 100~Hz repetition rate. The measurements show sub-ps stability already after 300~s of integration time, well below the performance specified for the ELT optical link.

ELT operation relies on existing SLR stations with a stable timing infrastructure connected to atomic clocks. To reach the specified link stability and accuracy for time transfer experiments, the optical ground stations need to be characterized and their delays measured. A first calibration campaign of the European SLR stations (Wettzell, Graz, Herstmonceux, Potsdam, and Zimmerwald) has already been completed in 2016 to demonstrate the feasibility of the ELT experiment~\cite{Prochazka2017}. SLR station delays were monitored up to 1 year, showing peak-to-peak variations at the 20~ps level. A new calibration campaign will be performed before the ACES mission launch. 

There is no major limitation to the number of SLR station that can participate to the ELT experiment, provided that they comply with the ISS safety standards for laser emission.

\section{Conclusions}
The ACES payload is currently completing the qualification tests. The PHARAO and SHM clocks, the frequency comparator FCDP, the onboard GNSS receiver, the power distribution unit, and the onboard computer are all integrated in the ACES payload, and currently tested at system level to measure the performance of the servo-loops and of the ACES clock signal. The MWL flight electronics is continuing its qualification tests before final integration. MWL calibrations will be soon established together with the characterization of the link delays for time transfer experiments. After completion of the MWL calibration campaign, the flight hardware will enter the final performance tests where, in end-to-end configuration with the MWL ground terminal electronics, the link performance will be established under a signal dynamics representative of the ISS orbit. After the integration of MWL, the ACES payload will be in its final configuration. The clocks and the links will then be commanded and controlled in a flight representative operational environment for the final acceptance tests before flight.
ACES will be ready for flight in the last quarter of 2024.

\ack
The authors wish to acknowledge the contribution of the ACES science team members, the ACES project team at ESA, the ACES teams at Airbus Defence and Space, Timetech, Safran, and the PHARAO team at SYRTE and CNES.

\end{document}